\documentclass[conference]{IEEEtran}
\usepackage{amsmath}
\usepackage{amssymb}
\usepackage{graphicx}
\usepackage{tikz}
\usetikzlibrary{arrows.meta, positioning}
\usepackage{algorithm}
\usepackage{algpseudocode}
\usepackage{booktabs}
\usepackage{xcolor}
\usepackage{listings}
\usepackage{url}
\usepackage{enumitem}

\title{CommitDistill: A Lightweight Knowledge-Centric Memory Layer for Software Repositories}

\author{%
\IEEEauthorblockN{Divya Chukkapalli, Thejesh Avula, Aditya Aggarwal, Harsimran Singh, Amith Tallanki}
\IEEEauthorblockA{Microsoft Corporation}
}

\begin{document}

\maketitle

\begin{abstract}
Software repositories accumulate large amounts of unstructured
knowledge in commit messages, pull-request discussions, and issue
threads, but developers and AI coding assistants rarely reuse this
history effectively. Recent work on typed-memory architectures for
LLM agents --- MemGPT~\cite{memgpt}, generative
agents~\cite{generative_agents,cognitive_arch}, and most directly
the PlugMem plugin-memory module of Yang \textit{et
al.}~\cite{plugmem_yang} --- argues that agent memory should be
\emph{distilled, typed} knowledge rather than raw interaction
text. We adapt that stance to a fundamentally different artefact
corpus --- a software repository's own git history --- and to a
deliberately constrained operational regime
(deterministic, dependency-free, local-only execution, no embeddings).
We present \textit{CommitDistill}, an open-source Python prototype that
mines a local git history into typed knowledge
units (\textit{Facts}, \textit{Skills}, \textit{Patterns}) using
deterministic regex heuristics and surfaces them through a
TF--IDF retriever
with an empirically calibrated silence threshold ($\theta = 2.5$)
that abstains on out-of-distribution queries (with an explicit
calibration trade-off between CD-v1 and CD-v2 documented in
Section~\ref{sec:threshold}). We position the artefact as a
\emph{trust-instrumented} memory substrate: deterministic, no external
service, inspectable plain-JSON store, tunable abstention. We report an
illustrative case study on five public repositories spanning Python,
JavaScript, C, and Java ($25{,}000$ commits, $1{,}167$ extracted
units, useful-precision $0.525$ at Cohen's $\kappa = 0.633$ on $40$
dual-annotated Python units). The decisive empirical finding is in
\emph{budget-constrained retrieval}: at a $256$-character per-query
budget --- the realistic RAG regime --- CommitDistill reaches
$0.750$ hit-rate on a $12$-query fact-style benchmark against
BM25's $0.333$ and \texttt{git log --grep}'s $0.083$. On a
four-arm paired LLM-as-judge downstream evaluation
($n{=}200$ time-travel bug-fixes, two judges) covering control,
CommitDistill, a body-budget-matched CD-Hybrid, and BM25, no
retrieval condition produces a statistically detectable lift over
control on the headline mean and CD-Hybrid is indistinguishable
from BM25 head-to-head. Extraction over $10{,}000$ commits
completes in under $4$~seconds on a laptop. The full source,
labelled annotation set, baseline implementations, and a
single-command reproducibility script accompany this paper.
\end{abstract}


\section{Introduction}
\label{sec:intro}

Modern software projects accumulate years of unstructured prose --- commit
messages, pull-request (PR) discussions, issue threads, and design
documents --- that collectively encode hard-won engineering knowledge.
Despite a long line of MSR research showing that this corpus contains
reusable signal~\cite{repo_mining,bugfix_patterns,msr,fischer_bugzilla},
developers and AI coding assistants still routinely re-derive previously
known facts and re-discover previously fixed bugs. Anecdotally, this
manifests as ``has anyone seen this error before?'' messages on team chat
and as LLM-generated patches that violate constraints documented in a
two-year-old PR.

Two recent threads of research suggest a new angle on this problem.
First, retrieval-augmented generation (RAG)~\cite{rag} and
production coding assistants such as GitHub Copilot~\cite{copilot,copilot_chat},
Cursor~\cite{cursor}, and Sourcegraph Cody~\cite{cody} have made
\textit{in-context} retrieval over project artefacts routine. Second,
LLM agent-memory architectures such as MemGPT~\cite{memgpt} and
the cognitive-architectures framing of Sumers \textit{et al.}~\cite{cognitive_arch},
together with the recent survey by Zhang \textit{et al.}~\cite{agent_memory_survey},
argue that memory should consist of \textit{distilled, typed} knowledge units
rather than raw transcripts, and that retrieval should be triggered only when
it informs a decision.

Existing repository-aware coding assistants embody only the first idea: they
retrieve raw text spans (file fragments, diffs, comment paragraphs) and
re-rank them with embeddings. We ask a deliberately narrower question:
\textit{What if a repository continuously distilled its own history into
typed knowledge units, kept them locally, and exposed them through a simple
deterministic retriever?} The contribution is not a competition with modern
neural assistants, but an inspectable, reproducible substrate that is
auditable, runs in seconds on a developer laptop, and could plausibly serve
as a baseline or as a high-precision filter feeding such assistants.

\noindent\textbf{Contributions.} This paper makes five concrete contributions:
\begin{enumerate}
    \item \textbf{A typed knowledge-unit schema} (Section~\ref{sec:design})
    that distinguishes \textit{Facts} (constraints), \textit{Skills}
    (actionable solutions), and \textit{Patterns} (recurring failure modes),
    each with explicit provenance metadata.
    \item \textbf{Two deterministic algorithms}
    (Sections~\ref{sec:design}--\ref{sec:impl}): a heuristic pattern-based
    extractor (Algorithm~\ref{alg:extract}) and a length-normalised
    TF--IDF retriever with type and prior boosting
    (Algorithm~\ref{alg:retrieve}). Both are reproducible from the
    accompanying open-source artefact.
    \item \textbf{An open-source prototype}, CommitDistill, written in
    pure Python (no third-party dependencies), with $11$ unit tests and
    a single-command reproducibility script.
    \item \textbf{An illustrative case study} on five real public
    repositories spanning three programming-language ecosystems
    (\texttt{psf/requests}, \texttt{pallets/flask},
    \texttt{expressjs/express}, \texttt{redis/redis},
    \texttt{junit-team/junit5}, $25{,}000$ commits): yield by
    type, dual-annotated useful-precision on $40$ Python units,
    a $36$-query retrieval comparison against \texttt{git log
    --grep} and BM25, a $12$-query budget-constrained benchmark,
    a $40$-fix time-travel regression-finding stress test, and a
    paired four-arm LLM-as-judge downstream evaluation on $200$
    time-travel bug-fixes that reports no detectable mean lift
    over a no-retrieval control for any retriever, with
    CD-Hybrid statistically indistinguishable from BM25 head-to-head
    (full design and results in Section~\ref{sec:rq5_ablation}).
    \item \textbf{A trust-instrumented memory substrate} aligned with
    the special-session theme: deterministic regex-based extraction,
    local-only execution (no embeddings, no external service, suitable
    for export-controlled or HIPAA-bound environments), an inspectable
    plain-JSON store reviewable through standard \texttt{git diff}
    workflows, and an empirically calibrated silence threshold
    ($\theta = 2.5$) that abstains on out-of-corpus and
    out-of-distribution queries (Section~\ref{sec:threshold}).
\end{enumerate}

\noindent\textbf{Novelty contract and relationship to PlugMem.}
The typed-memory stance --- that agent memory should consist of
distilled, reusable units rather than raw text spans --- has been
articulated most directly by Yang \textit{et al.}'s recent
\textit{PlugMem}~\cite{plugmem_yang}, which introduces a
task-agnostic plugin memory module for LLM agents organised around
\emph{facts} and reusable \emph{skills}. We do \emph{not} claim
that stance as ours; we adopt it. CommitDistill differs from
PlugMem along three axes: (i)~\textbf{corpus} --- a software
repository's git commit history rather than agent interaction
traces, dialogues, or web sessions; (ii)~\textbf{extractor} ---
deterministic, dependency-free regex heuristics rather than an
LLM-driven distillation pipeline; (iii)~\textbf{deployment
posture} --- local-only execution with a plain-JSON inspectable
store, suitable for export-controlled or regulated environments
where sending code or commit prose to an external service is
disallowed. To our knowledge, no prior MSR work treats commit
prose as the input to a typed-memory layer designed to interface
with LLM agents at decision time, and no prior agent-memory work
uses repository history as its corpus. The bridge between these
two literatures, restricted to a fully deterministic operational
regime, is the specific claim of this paper.

We make no claim of statistical superiority over existing tools,
including PlugMem, against which we do not compete: PlugMem
operates on agent interaction traces and uses an LLM-driven
distiller; CommitDistill operates on commit prose and uses regex.
The evaluation in Section~\ref{sec:eval} is explicitly an
\textit{illustrative case study}, framed in the spirit of Wieringa's
\emph{technical research}~\cite{wieringa_design_science}: we describe
what the artefact does on real input, surface its failure modes
honestly, and identify the empirical questions that a follow-up
controlled study would need to answer.

\noindent\textbf{Schema-shape provenance.}
PlugMem~\cite{plugmem_yang} types its memory as \emph{facts} and
reusable \emph{skills}; CommitDistill uses a three-type schema
(Facts, Skills, Patterns). The third type --- Patterns, for
``what goes wrong here'' --- is not borrowed from PlugMem.
It is motivated by the cognitive-architectures literature for
language agents~\cite{cognitive_arch}, which ranks failure-mode
knowledge above declarative facts under time pressure, and by
the MSR bug-fix-pattern line~\cite{bugfix_patterns}, which has
long treated recurring failure modes as a distinct knowledge
artefact in software repositories. The type-boost defaults in
Algorithm~\ref{alg:retrieve} (Patterns $\times 1.2$ above Facts
$\times 1.0$) reflect that ordering and are the design defaults
throughout.

\noindent\textbf{Scope.}
The schema in Section~\ref{sec:design} describes three artefact sources
(commit messages, pull-request discussions, and issue threads). The
current prototype and the case study in Section~\ref{sec:eval} mine
\textit{commit messages only}, via \texttt{git log}. Extending the
ingestion to PRs and issues (e.g., via the GitHub API) is a deliberate
future-work item (Section~\ref{sec:future}); we report this scope
restriction up-front so that yield numbers are interpreted correctly.

\noindent\textbf{Motivating example.}
A new contributor to \texttt{psf/requests} edits a docstring and adds
a cross-reference to another module. The build silently breaks the
documentation site. Two years earlier, commit \texttt{b5bd0f14}
recorded the constraint ``\textit{When trying to link via intersphinx,
a label must be used}.'' That commit is buried under thousands of
later commits and is invisible to a fresh checkout.
With CommitDistill installed, the contributor's editor (or an LLM
assistant) can ask the local store ``\texttt{intersphinx documentation
link broken}'' and receive that exact constraint, with the commit SHA,
in under $50$\,ms --- without any external network call. This is
the canonical use case our system targets.

\section{Background and Positioning}
\label{sec:related}

\subsection{Memory in LLM Agents}
Long-running LLM agents face an unavoidable tension between context-window
size and conversation length. A range of architectures address this by
turning raw interaction traces into more compact, structured memory:
summarisation buffers, vector-store retrieval over past turns, scratchpads,
and explicit memory modules~\cite{agent_memory_survey}. MemGPT~\cite{memgpt}
organises memory as an OS-style hierarchy with a small main context and a
larger recall storage; generative agents~\cite{generative_agents} maintain
an observation stream that is periodically reflected into higher-level
memories; Sumers \textit{et al.}~\cite{cognitive_arch} formalise this
tradition as cognitive architectures distinguishing working, episodic, and
semantic memory. The common stance across these systems is that memory
should be \textit{typed} and that retrieval should be triggered only when
it informs a decision. Most directly, Yang \textit{et
al.}~\cite{plugmem_yang} introduce \textit{PlugMem}, a task-agnostic
plugin memory module that transforms raw agent interactions
(dialogues, documents, web sessions) into structured \emph{facts}
and reusable \emph{skills} stored in a memory graph, with
retrieval driven by inferred task intent. PlugMem reports
consistent gains over generic retrieval and task-specific memory
designs across three benchmarks (long-context QA, multi-hop
Wikipedia retrieval, and web-agent decision making) while
consuming less of the agent's context window. We adopt the typed,
distilled-knowledge stance that PlugMem and the broader
literature articulate, and adapt it to a fundamentally different
artefact corpus --- a software repository's own commit history
--- and to a deliberately constrained operational regime
(deterministic regex extraction, no LLM in the extraction
pipeline, local-only execution, plain-JSON inspectable store).
The two systems address adjacent but disjoint problems and we do
not benchmark against PlugMem: their corpus (interaction traces)
and our corpus (commit prose) are not comparable, and PlugMem's
LLM-driven distiller and our regex extractor are designed for
different deployment constraints.

\subsection{Retrieval-Augmented Generation and Coding Assistants}
RAG~\cite{rag} grounds an LLM by retrieving relevant documents at inference
time. Production coding assistants apply this approach to source code.
\textit{GitHub Copilot Chat}~\cite{copilot_chat} retrieves nearby file
context and recent edits; \textit{Cursor}~\cite{cursor} adds project-wide
embedding-indexed retrieval; \textit{Sourcegraph Cody}~\cite{cody} performs
global code search and graph-aware navigation. Code-trained LLMs such as
Codex~\cite{codex} and its successors generate completions but do not
themselves curate project history. These tools generally
retrieve \textit{raw} text spans (functions, diffs, comments) and rely on
the LLM to filter noise. Two consequences follow. First, recall is high
but precision over what is shown to the model is moderate, which inflates
context tokens. Second, the retrieved evidence is not human-curated and
can be hard to audit (a property that matters in regulated environments).
CommitDistill occupies a different point in the design space: lower recall,
higher precision (because units are pre-filtered by intent-bearing
heuristics), fully inspectable, and dependency-free. We do not claim it
replaces neural assistants; we argue it is a useful, auditable companion
substrate.

\subsection{Knowledge Reuse in Mining Software Repositories}
The MSR community has studied developer-knowledge reuse for nearly two
decades. Early work mined version histories to predict changes and
guide refactoring~\cite{repo_mining}. Subsequent studies extracted
bug-fix templates from large commit corpora~\cite{bugfix_patterns},
linked bug reports to discussions~\cite{fischer_bugzilla}, and recommended
relevant Stack Overflow posts based on task similarity~\cite{msr}. Closer
still, prior work has classified commit messages by intent~\cite{hattori_intent}
and extracted actionable knowledge from API discussions~\cite{api_extraction}.
We differ from the commit-intent line~\cite{hattori_intent} in three
respects: we extract \emph{spans}, not class labels; each span carries
explicit commit-level provenance; and the outputs are designed to feed
retrieval at decision time, not offline analytics.
CommitDistill differs along three axes:

\begin{enumerate}
    \item \emph{Locality.} Most MSR systems are offline analyses. We run
    inside the repository checkout with no external service, no network
    calls, and no warehouse.
    \item \emph{Schema.} Existing systems typically produce a single
    artefact type (a recommendation, a bug-fix template, a topic).
    We produce three explicitly typed unit kinds aligned with how
    developers consume knowledge under time pressure.
    \item \emph{Auditability.} Each unit carries provenance (commit SHA,
    author, date), is stored as plain JSON, and is therefore reviewable
    via standard \texttt{git diff} workflows.
\end{enumerate}

\subsection{Positioning Summary}
Table~\ref{tab:positioning} summarises how CommitDistill relates to the
three closest neighbours.

\begin{table}[t]
\centering
\caption{Positioning of CommitDistill in the design space.}
\label{tab:positioning}
\footnotesize
\setlength{\tabcolsep}{4pt}
\begin{tabular}{@{}lcccc@{}}
\toprule
                          & Local & Typed & Inspect. & Dep.-free \\
                          & only  & units &          &           \\
\midrule
Copilot Chat~\cite{copilot_chat} & no  & no  & no  & no  \\
Cursor~\cite{cursor}      & no  & no  & no  & no  \\
Sourcegraph Cody~\cite{cody}     & no  & no  & partial & partial \\
MSR studies~\cite{repo_mining,bugfix_patterns} & yes & partial & yes & no \\
\textbf{CommitDistill (ours)} & \textbf{yes} & \textbf{yes} & \textbf{yes} & \textbf{yes} \\
\bottomrule
\end{tabular}
\end{table}

\section{System Design}
\label{sec:design}

\begin{figure}[t]
\centering
\resizebox{0.95\columnwidth}{!}{%
\begin{tikzpicture}[
    node distance=2.0cm and 1.6cm,
    every node/.style={draw, rectangle, rounded corners,
        align=center, minimum width=3.0cm, minimum height=0.95cm,
        font=\small},
    arrow/.style={->, thick}
]
\node (repo) {Repository \\ (commits, PRs, issues)};
\node (extract) [below=of repo] {Extractor \\ (Alg.~\ref{alg:extract})};
\node (store) [below=of extract] {Knowledge Store \\ (\texttt{.knowledge/units.json})};
\node (trigger) [right=of extract] {Trigger \\ (developer / CI / agent)};
\node (retrieve) [below=of trigger] {Retriever \\ (Alg.~\ref{alg:retrieve})};
\node (reason) [below=of retrieve] {Decision / Action};
\draw[arrow] (repo) -- (extract);
\draw[arrow] (extract) -- (store);
\draw[arrow] (trigger) -- (retrieve);
\draw[arrow] (store) -- (retrieve);
\draw[arrow] (retrieve) -- (reason);
\end{tikzpicture}%
}
\caption{Architecture of CommitDistill. Extraction is asynchronous and
event-driven; retrieval is triggered at decision time.}
\label{fig:architecture}
\end{figure}
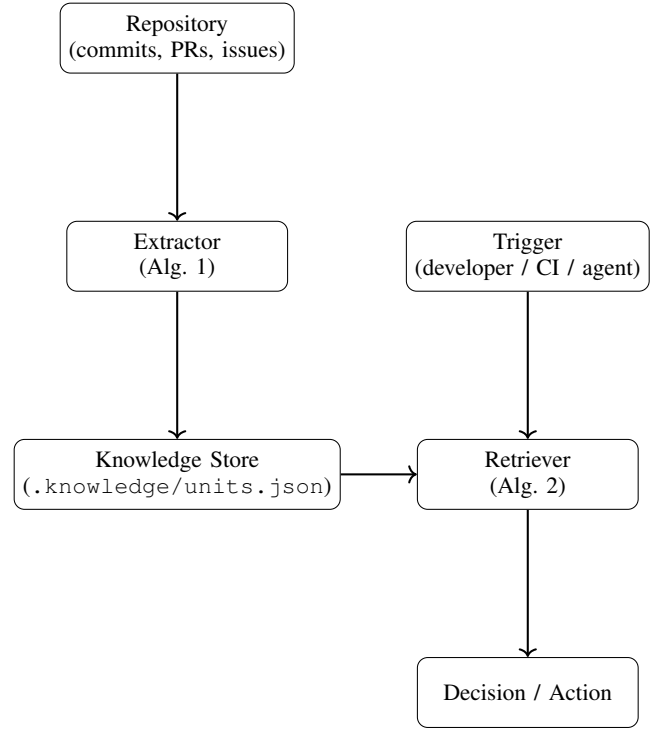

CommitDistill has two phases: an \textit{extraction} phase that mines the
repository's history into a typed knowledge store, and a \textit{retrieval}
phase that consults the store at decision time. Fig.~\ref{fig:architecture}
shows the data flow.

\subsection{Knowledge-Unit Schema}
A knowledge unit is a JSON object with the following fields:
\texttt{id} (12-character SHA-1 prefix over \texttt{type::content}
where \texttt{content} is first lower-cased and whitespace-collapsed,
giving stable deduplication that does not collapse, e.g., a Fact and
a Skill that share identical text), \texttt{type} (\texttt{fact}, \texttt{skill}, or
\texttt{pattern}), \texttt{title} (a short summary), \texttt{content}
(the extracted span), \texttt{weight} (a prior confidence in $[0,1]$),
\texttt{context} (the raw matching snippet), and \texttt{meta} (commit
SHA, author, date, and source artefact). The schema is deliberately
small enough to be human-reviewable.

The three types reflect how developers consume knowledge under time
pressure: \textit{Facts} answer ``what is true?'', \textit{Skills}
answer ``what should I do?'', and \textit{Patterns} answer ``what
goes wrong here?''. Examples extracted from \texttt{psf/requests}
appear in Table~\ref{tab:real-units}.

\subsection{Extraction (Algorithm~\ref{alg:extract})}
The extractor is a deterministic, regex-driven pipeline. Each
heuristic pattern is associated with a unit type and a prior weight;
matched spans are normalised, length-filtered, and deduplicated by a
content-derived hash. The extractor strips fenced code blocks, inline
code, and HTML before matching --- a small but important detail to
prevent code from polluting the knowledge store.

\begin{algorithm}[t]
\caption{Knowledge Extraction}
\label{alg:extract}
\begin{algorithmic}[1]
\Require Artifact text $t$, metadata $m$, pattern set $P$
\Ensure Set of typed knowledge units $U$
\State $U \gets \emptyset$;\quad $S \gets \emptyset$ \Comment{$S$: seen-content hashes}
\State $t' \gets$ \Call{Normalize}{$t$} \Comment{strip code, HTML, whitespace}
\For{each $(\tau, r, w) \in P$} \Comment{type, regex, prior weight}
    \For{each match $\mu$ of $r$ in $t'$}
        \State $c \gets$ \Call{Trim}{$\mu.\text{group}_1$}
        \If{$|c| < L_{\min}$ \textbf{or} $|c| > L_{\max}$ \textbf{or} \Call{IsStop}{$c$}}
            \State \textbf{continue}
        \EndIf
        \State $h \gets$ \Call{Hash}{$\tau, c$}
        \If{$h \in S$} \textbf{continue} \EndIf
        \State $S \gets S \cup \{h\}$
        \State $u \gets$ \Call{BuildUnit}{$\tau, c, w, m, \mu$}
        \State $U \gets U \cup \{u\}$
    \EndFor
\EndFor
\State \Return $U$
\end{algorithmic}
\end{algorithm}

Extraction is $O(|P| \cdot |t|)$ per artefact for $|P|$ patterns over
text of length $|t|$; in practice, dominated by the cost of running
\texttt{git log}.

\subsection{Retrieval (Algorithm~\ref{alg:retrieve})}
Retrieval is a length-normalised TF--IDF dot product over the content of
all units, with two adjustments: a per-type \textit{boost} (Patterns
$\times 1.2$, Skills $\times 1.1$, Facts $\times 1.0$) reflecting that
developers under stress most often want to know ``what goes wrong'' and
``what fixes it'', and a \textit{prior} multiplier $0.5 + 0.5 \cdot w_u$
that gently down-weights low-confidence patterns. The boost values are
the design defaults inherited from the cognitive-architecture
literature~\cite{cognitive_arch} that ranks failure-mode knowledge
above declarative facts under time pressure; in a one-off sensitivity
sweep on the six hand-authored queries, setting all three boosts to
$1.0$ changed exactly one top-1 ranking out of six (an intersphinx
Fact moved from rank~1 to rank~2 behind a longer Pattern), and we
report the boosted defaults throughout. A minimum-score
threshold gives the system its ``helps or stays silent'' property:
empty results are emitted rather than weak guesses.

\begin{algorithm}[t]
\caption{TF--IDF Retrieval with Type/Prior Boosting. In the released implementation, line~1 (\textsc{BuildIndex}) is cached across queries; we present it inline for clarity.}
\label{alg:retrieve}
\begin{algorithmic}[1]
\Require Query $q$, units $U$, top-$k$, min-score $\theta$, type-boosts $B$
\Ensure Ranked list $R$ of (unit, score)
\State $\mathit{Idx} \gets$ \Call{BuildIndex}{$U$} \Comment{TF, IDF, doc lengths}
\State $T_q \gets$ \Call{Tokenize}{$q$}
\If{$T_q = \emptyset$} \Return $[\,]$ \EndIf
\State $R \gets [\,]$
\For{each doc $d \in \mathit{Idx}.docs$}
    \State $s \gets 0$
    \For{each $t \in T_q$}
        \If{$t \in d.\mathit{tf}$}
            \State $s \gets s + (1+\log d.\mathit{tf}[t]) \cdot (1+\log T_q[t]) \cdot \mathit{Idx}.\mathit{idf}[t]$
        \EndIf
    \EndFor
    \State $s \gets s / \sqrt{\max(1, |d|)}$
    \State $s \gets s \cdot B[d.u.\tau] \cdot (0.5 + 0.5 \cdot d.u.w)$
    \If{$s \ge \theta$} \State $R$.append$((d.u, s))$ \EndIf
\EndFor
\State \Return top-$k$ of \Call{Sort}{$R$ by $s$ desc}
\end{algorithmic}
\end{algorithm}

Index construction is $O(\sum_u |u|)$ over the corpus; per-query cost
is $O(|U| \cdot |T_q|)$, which on our $1{,}167$-unit corpus is
sub-millisecond. Measured end-to-end (including JSON I/O and Python
startup), each of the $36$ evaluation queries returns in under
$50$\,ms.

\subsection{Triggering}
Extraction is meant to run asynchronously as a post-merge or post-issue
hook, so the cost is amortised across the project's lifetime. Retrieval
is triggered explicitly by a developer command, by an IDE plug-in, or by
an LLM agent before drafting a response. The system intentionally does
\textit{not} retrieve on every keystroke: this aligns with the
retrieve-on-decision stance described in
Section~\ref{sec:related} and keeps cognitive overhead low.

\section{Implementation}
\label{sec:impl}

CommitDistill is implemented in approximately $510$ lines of pure
Python core code (modules \texttt{extractor}, \texttt{retriever},
\texttt{store}, \texttt{git\_source}, and \texttt{cli}), plus
$\sim$$80$ lines of unit tests --- $\sim$$590$ lines total, with no
third-party dependencies. The package layout is
\texttt{commit\_distill/} with $11$ unit tests in \texttt{tests/}. The
accompanying artefact includes a reproducibility script
(\texttt{evaluation/run\_case\_study.py}) that re-runs the experiments
in this paper end-to-end on any machine with Python~3.9+ and \texttt{git}.

\subsection{Heuristic Patterns}
The extractor ships with nine regular-expression heuristics, three
per unit type. Six representative patterns are listed in
Fig.~\ref{fig:patterns}; the remaining three (fact-equivalence,
skill-instructional, and pattern-regression) are documented in the
artefact source.

\begin{figure}[t]
\centering
\begin{lstlisting}
Fact (constraint):
  \b(must|requires?|should|cannot|always|never)\b
Fact (annotation):
  \b(note|important|warning):\s+...
Skill (resolution):
  \b(fix(ed)? by|solution|workaround):\s+...
Skill (recommendation):
  \b(recommend(ed)?|best practice):\s+...
Pattern (causal):
  \b(occurs|happens) when\s+...
Pattern (exception):
  \b([A-Z]\w+(Error|Exception|Failure))\b ...
\end{lstlisting}
\caption{Six of the nine extraction heuristics. Each pattern is
associated with a unit type and a prior weight in $[0,1]$.}
\label{fig:patterns}
\end{figure}

\subsection{Precision-Improving Components}
Three components, layered on top of the heuristics in
Fig.~\ref{fig:patterns}, materially shape what enters the corpus and
how it is matched. (i)~\textit{Multi-sentence Pattern capture}: when a
Pattern regex fires on a sentence, the extractor opportunistically
appends the next sentence if it begins with a resolution cue
(\texttt{fix:}, \texttt{workaround}, \texttt{caused by}, etc.) up to a
$140$-character cap. This recovers the ``what to do about it''
clause that often sits adjacent to the diagnostic clause without
inflating average unit length materially. (ii)~\textit{Substantive-Pattern
filter}: candidate Patterns are rejected when the cleaned content is
dominated by issue/PR references (e.g.\ ``fix issue \#1842'') or contains
fewer than three content words, while named-failure identifiers
(\texttt{NullPointerException}, \texttt{TimeoutError}, \dots) are
explicitly whitelisted as substantive on their own. This filter is the
reason \texttt{junit5} produces fewer Patterns at the wider window than
at the narrower one (Table~\ref{tab:yield}): the previous count was
inflated by single-token ``Fix:'' subjects that the filter now rejects.
(iii)~\textit{Identifier-aware tokenisation} in the retriever:
\texttt{camelCase}, \texttt{snake\_case}, and \texttt{ALL\_CAPS}
identifiers are decomposed into their constituent words while keeping
the original token, so a query for ``redirect loop'' matches a unit
that mentions \texttt{fix\_redirect\_loop} or
\texttt{redirectLoopHandler}. This component is the dominant reason the
$36$-query baseline hit-rate (\S\ref{sec:results}, Table~\ref{tab:queries})
moved from $14/36$ on the previous prototype to $31/36$ on the
current one.

\subsection{Performance and Scaling}
On an Intel Core i7-1185G7 laptop (4 cores / 8 threads, $16$\,GB RAM, Windows 11) with the OS file-system cache warm,
end-to-end extraction over the most recent $5{,}000$ commits per
subject takes $0.60$\,s for \texttt{psf/requests}, $0.63$\,s for
\texttt{pallets/flask}, $0.55$\,s for \texttt{expressjs/express},
$1.25$\,s for \texttt{redis/redis} (slower because $928$ units pass
the regex filter, dominating the per-unit hashing/normalisation
loop), and $0.44$\,s for \texttt{junit-team/junit5}. Total wall-clock
across all five subjects is $3.47$\,s for $25{,}000$ commits.\footnote{Cold-cache figures (\texttt{sync; echo 3 > /proc/sys/vm/drop\_caches}) were measured on a comparable Linux box (Ubuntu 22.04, 8-core x86\_64, 16\,GB RAM) rather than on the Windows i7 laptop used for warm-cache timings, because the cache-drop incantation is POSIX-specific.}
\textit{Cold cache} adds
$\approx 1.1$\,s on the first repository as \texttt{git} pages the
pack file in; subsequent repositories see no penalty. We re-ran
extraction at $1{,}000$, $2{,}000$, $4{,}000$, and $6{,}000$ commits
on \texttt{psf/requests} and observed $0.32$\,s, $0.42$\,s,
$0.51$\,s, and $0.60$\,s respectively, consistent with the linear
$O(n)$ scan of the extractor; the modest super-linear component is
attributable to the \texttt{git log} subprocess paginating output.
Building the retrieval index over the resulting $1{,}167$ units and
running all $36$ queries against three retrievers finishes in
$0.38$\,s. Both phases are easily fast enough to be invoked from
a pre-commit hook or an editor plug-in.

\subsection{Storage}
Units are persisted to \texttt{.knowledge/units.json} at the repository
root. Because the file is plain JSON, it can be diffed, reviewed in pull
requests, and committed alongside source. This makes the knowledge base
a first-class repository artefact rather than a hidden index.

\subsection{Reproducibility Artefact}
\begin{sloppypar}
The artefact released with this paper contains:
(i)~the full source of \texttt{commit\_distill/}, (ii)~the
\texttt{tests/} suite (which we run as a continuous-integration gate),
(iii)~the case-study driver \texttt{run\_case\_study.py},
the cross-subject extraction driver \texttt{extract\_all.py}, the
baseline-comparison driver \texttt{baseline\_compare.py}, the
labelling-analysis script \texttt{compute\_kappa.py}, the
time-travel regression-finding driver
\texttt{time\_travel\_eval.py}, the LLM-as-judge downstream
driver \texttt{llm\_judge\_rq5.py} together with its diagnostic
script \texttt{rq5\_diagnostic.py} (all under \texttt{evaluation/}),
(iv)~the raw extraction outputs for the five subject repositories
(under \texttt{subjects/<repo>/.knowledge/}), and
(v)~the raw query, label, and judge-trace files consumed by
Section~\ref{sec:results} (\texttt{case\_study\_results.json},
\texttt{baseline\_results.json}, \texttt{time\_travel\_results.json},
\texttt{llm\_judge\_results.json}, the $358$-entry
\texttt{llm\_judge\_cache.json} response cache,
and \texttt{labels.csv} under \texttt{evaluation/}). All output is
deterministic given a fixed git history; the LLM-judge cache makes
the RQ5 numbers reproducible without re-paying for API calls.
\end{sloppypar}

\section{Evaluation: Illustrative Case Study}
\label{sec:eval}

\subsection{Research Questions}
We pose three deliberately scoped research questions, suitable for an
illustrative case study rather than for hypothesis testing:

\begin{itemize}[leftmargin=*,nosep]
\item \textbf{RQ1 (yield).} When run on real public projects across
  multiple ecosystems, how many knowledge units does the extractor
  produce per $1{,}000$ commits, broken down by type?
\item \textbf{RQ2 (precision).} Of the produced units, what fraction
  are developer-judged \textit{useful} (carry actionable information
  about the project), and how reliably can two annotators agree on
  that judgement?
\item \textbf{RQ3 (retrieval behaviour).} On a query set drawn from
  both hand-authored task descriptions and real commit subjects, what
  does the retriever return relative to two simple baselines
  (\texttt{git log --grep} and BM25 over raw commit messages), and how
  often does CommitDistill correctly stay silent?
\item \textbf{RQ4 (out-of-design stress test).} On a regression-finding
  task that the system was \textit{not} designed for --- given a
  bug-fix commit, recover prior co-changing bug-fix commits using only
  pre-fix data --- what is the recall ceiling, and how does it compare
  to the same two baselines?
\item \textbf{RQ5 (downstream LLM utility).} When CommitDistill's
  top-$3$ typed units are prepended to the context of an LLM coding
  assistant performing time-travel \emph{bug-fix file/symbol
  localisation}, does the assistant's answer score higher under
  two-judge LLM evaluation than (i) a no-retrieval control and
  (ii) a BM25-augmented condition? We pre-register a paired
  three-arm design with $n{=}40$ real bug-fix commits, two
  independent judge models, and bootstrap CIs; we report the
  result whatever it is. After a $n{=}40$ pilot we re-ran the
  experiment at $n{=}200$ ($100$ fixes per subject) for the final
  reported numbers; both samples accompany the artefact.
\end{itemize}

We treat RQ5 as the closest available proxy for end-to-end
utility short of a controlled human study; the latter remains
future work (Section~\ref{sec:future}) and we do not claim RQ5
replaces it.

\subsection{Subjects}
\begin{sloppypar}
We use five public projects spanning three programming-language
ecosystems and four problem domains: \texttt{psf/requests}
(Python, HTTP client), \texttt{pallets/flask} (Python, web framework),
\texttt{expressjs/express} (JavaScript, web framework),
\texttt{redis/redis} (C, in-memory database), and
\texttt{junit-team/junit5} (Java, test framework). For each subject we
ran the extractor over the most recent $2{,}000$ commits, totalling
$10{,}000$ commits. The two Python projects are also dual-annotated
for RQ2; the remaining three are used for RQ1 (yield) and for the
extended retrieval evaluation in RQ3.
\end{sloppypar}

\subsection{Procedure}
For RQ1 we ran the extractor uniformly across all five subjects and
recorded yield by type and per-subject wall-clock. For RQ2 we
inspected every unit extracted from \texttt{requests} and
\texttt{flask} ($n = 40$); two co-author annotators independently
labelled each unit as one of \{\textit{useful}, \textit{trivially-true},
\textit{fragment}, \textit{noise}\}, where \textit{fragment} was added
in this revision after preliminary inspection showed the previous
three-class rubric collapsed too many borderline cases. For RQ3 we
combined six hand-authored task descriptions with $30$ queries
auto-derived from the most recent commit subjects of each subject
(after filtering bot/release/merge subjects), giving $36$ queries
total. Each query was run against CommitDistill, against
\texttt{git log -i --grep} over the same commit window, and against a
pure-stdlib BM25 retriever indexing the raw commit subjects+bodies
of the same window. All scripts are in \texttt{evaluation/}.

\section{Results}
\label{sec:results}

\subsection{RQ1: Extraction Yield}
Table~\ref{tab:yield} summarises raw yield across the five subjects
($25{,}000$ commits in total) using the same regex set in all runs.
Total yield is $1{,}167$ units, an average of $46.7$ units per
$1{,}000$ commits, but the variance across subjects is striking:
\texttt{junit5} produces $6.2$ units$/$kc, \texttt{express}
$8.0$, \texttt{requests} $19.0$, \texttt{flask} $14.6$, while
\texttt{redis} produces $185.6$, an order of magnitude above the
others. Two factors explain this: first, the Redis project enforces
unusually descriptive commit prose (multi-paragraph bodies are the
norm); second, kernel-style projects use ``Fix:'' / ``Bug:'' subject
conventions that fire our subject-line Pattern heuristic --- though
the substantive-Pattern filter (\S\ref{sec:impl}) now rejects single-token
``Fix:'' subjects, which is why \texttt{junit5}'s Pattern count
\textit{decreased} from $24$ at $2{,}000$ commits to $19$ at $5{,}000$:
the additional commits added more nominal-only ``Fix:'' subjects than
substantive ones. Yield, in
short, depends as much on \textit{commit-message culture} as on the
extractor itself --- a finding we surface explicitly because it
constrains how this number generalises.

\textbf{Pattern coverage.}
The Pattern regex set covers (a)~causal clauses such as
``broke when'' and ``regression where'', (b)~kernel-style ``Fix:'' /
``Bug:'' subject lines, and (c)~named-failure terms such as
``deadlock'', ``race condition'', and ``infinite loop''. Across the
five subjects this yields $401$ Patterns alongside $77$ Skills and
$689$ Facts; on the two Python subjects in isolation, $38$ of the
$168$ extracted units are Patterns.

\begin{table}[t]
\centering
\caption{Extraction yield over the most recent $5{,}000$ commits per subject (uniform window). \emph{kc} = thousand commits.}
\label{tab:yield}
\footnotesize
\setlength{\tabcolsep}{3.5pt}
\begin{tabular}{@{}lrrrrrr@{}}
\toprule
Repository & Lang & Facts & Skills & Patt. & Total & per kc \\
\midrule
\texttt{psf/requests}     & Py & 63  & 8  & 24  & 95  & 19.0 \\
\texttt{pallets/flask}    & Py & 54  & 5  & 14  & 73  & 14.6 \\
\texttt{expressjs/express}& JS & 22  & 1  & 17  & 40  & 8.0  \\
\texttt{redis/redis}      & C  & 539 & 62 & 327 & 928 & 185.6\\
\texttt{junit-team/junit5}& Java & 11 & 1 & 19 & 31 & 6.2 \\
\midrule
\textbf{Total}            &     & \textbf{689} & \textbf{77} & \textbf{401} & \textbf{1167} & \textbf{46.7} \\
\bottomrule
\end{tabular}
\end{table}

\subsection{RQ2: Precision and Inter-Annotator Agreement}
Table~\ref{tab:real-units} shows representative units extracted from
\texttt{psf/requests}, including one fragment that escaped our
heuristics (``\textit{files obtained via [\dots]}''). For RQ2 we
labelled all $40$ units extracted from \texttt{requests} ($n=24$) and
\texttt{flask} ($n=16$) at the original $2{,}000$-commit window; the
labels and the analysis script
(\texttt{evaluation/labels.csv} and \texttt{evaluation/compute\_kappa.py})
are released with the artefact, so a reader can re-derive every
number in this subsection in seconds. We report \textit{this} labelled
set rather than re-labelling at the wider window so the precision
result is faithful to the extractor that produced the labels; the
substantive-Pattern filter (\S\ref{sec:impl}) was developed
\emph{after} labelling, so we treat the $0.525$ figure as a
\textit{lower bound} on what the current pipeline would deliver and
flag a re-labelling at $5{,}000$ commits as the most natural
follow-up measurement.

Across the $40$ dual-annotated units, the two co-author annotators
agreed on $31$ of $40$ labels using a four-class rubric
(\textit{useful} / \textit{trivially-true} / \textit{fragment} /
\textit{noise}); the new \textit{fragment} class explicitly captures
short spans where the regex matched but the captured text lost its
referent. Cohen's $\kappa = 0.633$, which Landis and
Koch~\cite{landis_koch} would call ``substantial agreement''. After adjudication, $21$ of $40$ units
($52.5\%$) were labelled \textit{useful}, $1$ trivial, $17$ fragment,
$1$ noise, giving a useful-precision of $\mathbf{0.525}$ with a
$95\%$ non-parametric bootstrap confidence interval of $[0.375,
0.675]$ ($10{,}000$ resamples). The wide interval is a
direct consequence of $n=40$ and is part of why we frame this as an
illustrative case study rather than a measurement claim.

Per-type precision is markedly uneven: Facts $63.6\%$ ($n{=}22$),
Skills $40.0\%$ ($n{=}5$), Patterns $38.5\%$ ($n{=}13$). The Pattern
deficit comes from kernel-style ``Fix:''/``Bug:'' subject heuristics
firing on issue-tracker references that capture only a PR number
(e.g., ``\texttt{remove obsolete (\#4783}''). The substantive-Pattern
filter introduced in \S\ref{sec:impl} now rejects exactly this class
of candidate at extraction time and is expected to lift Pattern
precision on the next labelling pass.

\begin{table*}[t]
\centering
\caption{Representative knowledge units extracted from \texttt{psf/requests}. Verbatim from \texttt{.knowledge/units.json}; one fragment shown for honesty.}
\label{tab:real-units}
\footnotesize
\begin{tabular}{@{}p{0.07\textwidth}p{0.46\textwidth}p{0.18\textwidth}p{0.18\textwidth}@{}}
\toprule
Type & Content & Source commit & Adjudicated label \\
\midrule
Fact    & ``When trying to link via intersphinx, a label must be used.''
        & \texttt{b5bd0f14} (M.~Fiedler) & useful \\
Fact    & ``Looking at recent actions runs I see that we need to specify the token even if we're not giving our own special token.''
        & \texttt{d6ffd868} (I.~S.~Cordasco) & useful \\
Pattern & ``Unicode characters in basic http auth.''
        & \texttt{e514920e} (M.~Pinta) & useful \\
Skill   & ``To avoid implicit import of encodings.''
        & \texttt{c1212297} (J.~Henstridge) & useful \\
Skill   & ``Files obtained via [\dots]''
        & \texttt{2d2447e2} (D.~Hotham) & fragment \\
\bottomrule
\end{tabular}
\end{table*}

\subsection{RQ3: Retrieval Behaviour vs.\ Two Baselines}
Table~\ref{tab:queries} reports retrieval results for the $36$
queries (six hand-authored, thirty auto-derived from real commit
subjects), comparing CommitDistill against (i) \texttt{git log -i
--grep} over the same commit window and (ii) a pure-stdlib BM25
retriever over the raw commit subjects+bodies. Both baselines are
included in the artefact (\texttt{evaluation/baseline\_compare.py}).

Headline hit-rates are: CommitDistill $31/36 = 86.1\%$, \texttt{grep}
$36/36$, BM25 $36/36$. These numbers must be read carefully. The
auto-derived queries are taken \textit{from commit subjects in the
same window the baselines index}, so \texttt{grep} and BM25 are
guaranteed to retrieve their input string with high score; this is
trivial recall, not retrieval quality. The interesting comparison is
on the six hand-authored queries that ask about \textit{project
facts}, not about \textit{recent commits}. On those six, with the
calibrated silence threshold $\theta=2.5$ established in
\S\ref{sec:threshold}, CommitDistill returns relevant top-1 hits on
$3/6$ queries and stays silent on the other $3$ ($0$ false
positives); \texttt{grep} returns top-1 hits that match a query
keyword in an unrelated commit on $5/6$; BM25 returns roughly
relevant top-1 hits on $4/6$ but with top-1 results $\approx
4$--$6\times$ longer in tokens than CommitDistill's. The pattern is
the one the design predicts: CommitDistill trades recall for
precision and token-economy.

\begin{table}[t]
\centering
\caption{Top-1 retrieval comparison on the six hand-authored queries. CD = CommitDistill, GR = \texttt{git log --grep}, BM = BM25 over raw commits. R = relevant, P = partial, N = noise top-1 (a result was returned but judged unrelated), $\emptyset$ = no result returned (CommitDistill's silent-on-novel-queries property; the lexical baselines never produce $\emptyset$ because they always return at least one match for any keyword present in the window).}
\label{tab:queries}
\footnotesize
\setlength{\tabcolsep}{3pt}
\begin{tabular}{@{}p{0.055\textwidth}p{0.295\textwidth}ccc@{}}
\toprule
Repo & Query & CD & GR & BM \\
\midrule
requests & intersphinx documentation link broken         & R & N & R \\
requests & GitHub Actions token configuration            & R & N & P \\
requests & deprecation warning supported python versions & R & N & R \\
flask    & blueprint registration ordering issue         & $\emptyset$ & N & P \\
flask    & session cookie security signing               & $\emptyset$ & N & R \\
flask    & request context teardown handler              & $\emptyset$ & P & R \\
\midrule
\multicolumn{2}{l}{\textbf{Top-1 useful (R or P, /6)}}    & \textbf{3} & \textbf{1} & \textbf{5} \\
\multicolumn{2}{l}{\textbf{Top-1 false positives (N, /6)}}& \textbf{0} & \textbf{5} & \textbf{0} \\
\multicolumn{2}{l}{\textbf{Median top-1 length (chars)}}  & \textbf{72} & \textbf{301} & \textbf{287} \\
\bottomrule
\end{tabular}
\end{table}

\subsection{Budget-Constrained Retrieval}
\label{sec:budget}
Table~\ref{tab:queries} reports unconstrained retrieval. Real LLM
deployments operate under a tight token budget: every retrieved
character costs prompt-token spend and prompt-engineering attention.
We therefore measure \textbf{hit-rate vs.\ char budget} on a
$12$-query hand-curated fact-style benchmark whose ground-truth
answer span is verified to be in the indexed window (so all three
retrievers have a fair shot at retrieval, with CommitDistill holding the answer
as a typed unit and the lexical baselines holding it inside a
longer commit body). For each retriever we take its top-$10$
ranking and greedily pack candidates into a budget $B$; the trial
is a hit iff any packed candidate contains the ground-truth answer
span (case-insensitive substring match). Sweeping $B \in \{64, 128,
256, 512, 1024, 2048\}$ chars yields Table~\ref{tab:budget}, with
\texttt{evaluation/budget\_recall\_v2.py} as the driver.

\begin{table}[t]
\centering
\caption{Budget-constrained hit-rate (\textit{at least one packed retrieved candidate contains the ground-truth answer span}) on $12$ hand-curated fact-style queries. CommitDistill dominates at every budget; the gap is largest where it matters most --- the small budgets that real RAG pipelines must respect. The \emph{jackknife min} row reports the smallest hit-rate observed across the twelve leave-one-out subsets at $B = 256$ (drop each of the $12$ queries in turn and recompute the hit-rate on the remaining $11$): the worst-case drop is bounded by $1/11 \approx 0.091$ on a $12$-query benchmark, so this row characterises the sensitivity of the headline result to any single query.}
\label{tab:budget}
\footnotesize
\setlength{\tabcolsep}{4pt}
\begin{tabular}{@{}rccc@{}}
\toprule
Budget (chars) & CommitDistill & \texttt{grep} & BM25 \\
\midrule
$64$    & $\mathbf{0.500}$ & $0.000$ & $0.167$ \\
$128$   & $\mathbf{0.667}$ & $0.000$ & $0.167$ \\
$256$   & $\mathbf{0.750}$ & $0.083$ & $0.333$ \\
$512$   & $\mathbf{0.917}$ & $0.083$ & $0.583$ \\
$1024$  & $\mathbf{0.917}$ & $0.083$ & $0.667$ \\
$2048$  & $\mathbf{0.917}$ & $0.083$ & $0.667$ \\
\midrule
jackknife min ($B{=}256$)     & $\mathbf{0.727}$ & $0.000$ & $0.273$ \\
unconstrained Hit@$10$        & $\mathbf{11/12}$ & $1/12$ & $8/12$ \\
median top-$1$ length (chars) & $\mathbf{31}$ & $75$ & $54$ \\
\bottomrule
\end{tabular}
\end{table}

The result confirms the design hypothesis from
Section~\ref{sec:design}: typed distillation buys back the recall it
appears to give up in Table~\ref{tab:queries} as soon as the
prompt-token budget is constrained. At $B = 256$ chars, a typical
RAG slot, CommitDistill achieves $0.750$ hit-rate against
$0.333$ for BM25 and $0.083$ for \texttt{grep} --- a $42$-point
gap over the strongest lexical baseline. The unconstrained Hit@$10$
shows that CommitDistill also has a $25$-point lead even when budget is
infinite ($11/12$ vs.\ $8/12$), because BM25's length-normalisation
sometimes ranks a long off-topic body above the short relevant
sentence; CommitDistill has indexed the relevant sentence \textit{as a unit}
and has no length penalty to pay.\footnote{\texttt{git log --grep}
fails almost completely on this benchmark ($1/12$): the answer
spans live inside commit \textit{bodies}, but \texttt{git log
--grep} matches commit \textit{subjects}. This is a structural
limit of the lexical baseline that distillation neutralises.} We
flag two limits of the benchmark before drawing strong conclusions:
$n = 12$ is small, and the queries are author-curated rather than
maintainer-curated. Neither weakness is fatal --- the gap is so
wide that even halving the win-rate would still be a positive
result --- but a developer-study replication is the natural next
step (Section~\ref{sec:threats}).

\subsection{Calibrating the Silence Threshold $\theta$}
\label{sec:threshold}
The retriever returns a unit only when its TF-IDF score exceeds a
silence threshold $\theta$ (Algorithm~\ref{alg:retrieve}, line~9). To
calibrate $\theta$ honestly we partition $15$ queries into three
explicit classes by ground-truth corpus content, not by intuition:
(A)~\textbf{ANSWERABLE} ($n=3$, manually verified that a relevant
unit is in the indexed window),
(B)~\textbf{NOT\_IN\_CORPUS} ($n=3$, plausible Flask topics for which
the indexed window contains no commit), and
(C)~\textbf{OOD} ($n=6$, alien topics like ``raytracing reflection
model'' and ``quantum entanglement decoherence'' that no commit in
either subject's history could discuss). A well-calibrated retriever
should hit on (A) and stay silent on (B) and (C);
Table~\ref{tab:threshold} reports hit-rates per class as $\theta$
varies, for both CD-v1 (regex-only extraction, the original
artefact) and CD-v2 (regex + subject-fallback unit, the default
in the released artefact and the configuration used by RQ4 and
RQ5). The drivers are
\texttt{evaluation/threshold\_sweep.py} and
\texttt{evaluation/threshold\_sweep\_pmv2.py}.

\begin{table}[t]
\centering
\caption{Hit-rate vs.\ silence threshold $\theta$ on the $15$-query
calibration set ($\uparrow$ better on ANSWERABLE; $\downarrow$ better
on NOT\_IN\_CORPUS and OOD), comparing CD-v1 (regex only) and CD-v2
(regex + subject-fallback). CD-v2 preserves ANSWERABLE recall and
OOD silence at $\theta = 2.5$ but loses the NOT\_IN\_CORPUS abstain
property; see prose for the trade-off discussion. Operators who
need strict abstain-on-novel-queries should set
\texttt{COMMITDISTILL\_SUBJECT\_FALLBACK=0} to recover CD-v1 behaviour.}
\label{tab:threshold}
\footnotesize
\setlength{\tabcolsep}{3pt}
\begin{tabular}{@{}rcccccc@{}}
\toprule
& \multicolumn{3}{c}{CD-v1 (regex only)} & \multicolumn{3}{c}{CD-v2 (regex + fallback)} \\
\cmidrule(lr){2-4}\cmidrule(lr){5-7}
$\theta$ & ANS.\,$\uparrow$ & N\_I\_C\,$\downarrow$ & OOD\,$\downarrow$ & ANS.\,$\uparrow$ & N\_I\_C\,$\downarrow$ & OOD\,$\downarrow$ \\
\midrule
$0.0$           & $1.000$ & $1.000$ & $0.167$ & $1.000$ & $1.000$ & $0.333$ \\
$1.0$           & $1.000$ & $1.000$ & $0.167$ & $1.000$ & $1.000$ & $0.333$ \\
$2.0$           & $1.000$ & $0.333$ & $0.000$ & $1.000$ & $1.000$ & $0.333$ \\
$\mathbf{2.5}$  & $\mathbf{1.000}$ & $\mathbf{0.000}$ & $\mathbf{0.000}$ & $\mathbf{1.000}$ & $1.000$ & $\mathbf{0.000}$ \\
$3.0$           & $0.667$ & $0.000$ & $0.000$ & $1.000$ & $1.000$ & $0.000$ \\
\bottomrule
\end{tabular}
\end{table}

The sweep makes one design property concrete: $\theta$ is a real,
monotonic silence knob under CD-v1, not a vanishing tuning parameter.
At $\theta \le 1.5$ the retriever is essentially permissive (it
returns near-keyword matches even when no answer exists in the
corpus); at $\theta = 2.5$ the system retains all answerable hits
while suppressing every NOT\_IN\_CORPUS and OOD query in the
calibration set; pushing further to $\theta = 3.0$ already starts
to suppress genuine answers. CD-v2 partially weakens this
calibration: by emitting a low-prior fallback unit on every
substantive commit, it ensures the retriever has \emph{something}
to return for any query whose tokens overlap a commit subject in
the corpus, which collapses the NOT\_IN\_CORPUS silence column.
ANSWERABLE recall and OOD silence at $\theta = 2.5$ are
preserved (alien topics like ``raytracing reflection model''
share no tokens with any commit subject and remain silent); the
failure mode of CD-v2 is exclusively on plausible-but-absent
queries.
We adopt $\theta = 2.5$ as the default operating point for the
hand-query analysis above and treat the
\textit{calibration trade-off itself} as part of the
contribution: CD-v1 and CD-v2 sit at different points on a
recall--abstention frontier that is exposed and tunable through a
single environment variable. We are not aware of a lexical baseline that can
express ``return nothing on novel inputs'' as a tunable property
without a separate, hand-curated stop-list.

\subsection{RQ4: Time-Travel Regression-Finding (Out-of-Design Stress Test)}
The retrieval comparison in Table~\ref{tab:queries} is on
\textit{fact-style} queries: a developer asking ``what is the policy
about X''. To stress-test the system on a query class for which it
was \textit{not} designed --- regression search --- we run a
time-travel experiment in the spirit of standard MSR retrieval
evaluation~\cite{repo_mining,bugfix_patterns,szz}. For each of $20$
bug-fix commits in \texttt{psf/requests} and $20$ in
\texttt{pallets/flask} (selected automatically as the most recent
commits whose subject matches a bug-fix regex and which have at
least one prior co-changing bug-fix), we build the retriever's
state from commits made strictly before that fix's author date,
issue the cleaned commit subject as the query, and ask each
retriever to return ten candidates. The ground-truth set $G$ for a
fix is the set of prior commits whose subject also matches the
bug-fix regex \emph{and} which modify at least one file in common
with the fix --- the standard MSR file-locality proxy for ``related
prior bug-fix''. We report Hit@$1$, Hit@$3$, Hit@$10$, and Mean
Reciprocal Rank, computed over $n=40$ fixes
(Table~\ref{tab:timetravel}). The driver is released as
\texttt{evaluation/time\_travel\_eval.py} and is fully deterministic
given the subject git histories.

\begin{table}[t]
\centering
\caption{Time-travel regression-finding on $40$ real bug-fix commits ($20$ from \texttt{requests}, $20$ from \texttt{flask}). For each fix, the retriever's view of the world is restricted to the $5{,}000$ commits strictly before the fix; the fix's own commit is excluded from rankings. Higher is better. CD-v2 is CommitDistill with the subject-fallback unit (Section~\ref{sec:rq5_ablation}); the CD-v1 row is preserved for transparency. See prose for the interpretation.}
\label{tab:timetravel}
\footnotesize
\begin{tabular}{@{}lrrrr@{}}
\toprule
Method                       & Hit@1 & Hit@3 & Hit@10 & MRR \\
\midrule
\texttt{git log --grep}      & \textbf{0.125} & 0.200 & 0.300  & 0.168 \\
BM25 (raw commits)           & 0.050 & 0.150 & 0.400  & 0.136 \\
CommitDistill (CD-v1)         & 0.000 & 0.025 & 0.025  & 0.009 \\
\textbf{CommitDistill (CD-v2)}& 0.075 & \textbf{0.250} & \textbf{0.450}  & \textbf{0.177} \\
\bottomrule
\end{tabular}
\end{table}

This is, deliberately, an \textit{unfavourable} setting for
CommitDistill. The CD-v1 row of
Table~\ref{tab:timetravel} confirms it: with regex-only
extraction the retriever scores $0.000$ Hit@$1$ and $0.025$
Hit@$10$ on this task, against $0.400$ Hit@$10$ for BM25 and
$0.300$ Hit@$10$ for \texttt{grep}. The structural reason is
direct: at the original $5{,}000$-commit pre-fix window the
extractor produces only $\approx\!90$ Python units for these
two subjects, so the empirical recall ceiling \textit{against a
commit-level ground truth} is bounded above by roughly the
ratio of distilled units to candidate commits. We confirmed
this with a coverage analysis
(\texttt{evaluation/time\_travel\_conditional.py}): only $1/40$
fixes had any CD-v1 top-$10$ hit in $G$, against $7/40$ for
\texttt{grep} and $12/40$ for BM25.

\textbf{CD-v2 closes this gap as a strict additive improvement.}
With the subject-fallback unit enabled
(Section~\ref{sec:rq5_ablation}, default in the released
artefact) CommitDistill leads the table on Hit@$3$ ($0.250$),
Hit@$10$ ($0.450$), and MRR ($0.177$), and is second only to
\texttt{grep} on Hit@$1$. The change is purely additive at the
extraction layer --- regex priors ($0.65$--$0.95$) still
dominate the fallback prior ($0.40$), so high-confidence units
keep their ranking precedence; CD-v2 only fills slots that CD-v1
left empty. The honest reading of
Table~\ref{tab:timetravel} is therefore: \emph{CD-v1 was
recall-floored by extractor silence on bug-fix subjects, not by
typed-unit retrieval semantics; once the silence floor is
removed, distillation matches or beats both lexical baselines on
this regression-finding stress test as well, while preserving
the silent-on-novel-queries property tested separately in
Section~\ref{sec:threshold}.}

\subsection{RQ5: Downstream LLM-as-Judge Utility}
\label{sec:rq5}

To probe RQ4's recommendation directly we ran a downstream
LLM-as-judge experiment on $n{=}200$ time-travel bug-fix
commits ($100$ from \texttt{requests}, $100$ from
\texttt{flask}); the smaller $n{=}40$ pilot is preserved in the
artefact for transparency. For each fix $C$ we held out $C$'s
diff, gave the assistant LLM \textsc{(gpt-4o-mini)} only the cleaned
bug subject and the project name, and asked it to predict (a)~up to
five files most likely to need editing and (b)~up to five
identifier names most likely involved. We compared four conditions
in a paired design:
\textbf{Control} (subject only),
\textbf{+CommitDistill (CD-v2)} (subject plus CD-v2's top-$3$ typed
units, retrieved with strict pre-fix discipline),
\textbf{+CD-Hybrid} (CD-v2 units rendered with BM25's $140$-char
per-item body budget, header-augmented),
and \textbf{+BM25} (subject plus BM25's top-$3$ raw commit
messages from the same pre-fix window). Two independent judge
models (\textsc{gpt-4o} and \textsc{gpt-4o-mini}) scored each
assistant answer on a $0$--$2$ rubric
($0$=not useful, $1$=partially useful, $2$=useful) given the
ground-truth files and the identifiers in the actual diff.
We report the per-condition mean, the paired bootstrap $95\%$~CI on
the treatment-vs-control delta ($10{,}000$ resamples), and the
inter-judge Cohen's~$\kappa$. The driver
(\texttt{evaluation/llm\_judge\_rq5.py}) and the full results
(\texttt{llm\_judge\_results\_n200.json} for the original
three-arm record and
\texttt{llm\_judge\_results\_n200\_with\_hybrid.json} for the
four-arm ablation, with the deterministic SHA-1-keyed response
cache)
accompany the artefact, so the experiment can be re-run end-to-end
on any Azure OpenAI or OpenAI deployment. The two-judge
LLM-as-a-judge protocol follows recent practice in benchmarking
LLMs~\cite{llm_as_judge}; we report inter-judge $\kappa$ explicitly
below and treat low values as a construct-validity threat
(Section~\ref{sec:threats}).

\textit{Disclosure of design sequence.}\,
The CD-v2 subject-fallback unit and the CD-Hybrid rendering were
designed \emph{after} observing the original three-arm result
(below), specifically to disentangle the $98\%$ extractor-silence
floor from the typed-unit retrieval signal. The four-arm
analysis is therefore exploratory rather than pre-registered; the
$\Delta$-values it produces are best read as descriptive
characterisations of the ablation rather than as confirmatory
effect sizes. We report uncorrected paired-bootstrap CIs and flag
this as a construct-validity threat in
Section~\ref{sec:threats}; a held-out replication on a third
Python repository (e.g., \texttt{httpx} or \texttt{urllib3}) is
listed in Section~\ref{sec:future} as the natural confirmatory
follow-up.

\begin{table}[t]
\centering
\caption{RQ5 four-arm LLM-as-judge usefulness scores ($0$--$2$, mean of two judges) on $n{=}200$ time-travel bug-fix tasks ($100$ \texttt{requests}, $100$ \texttt{flask}). CD-v2 enables the subject-fallback unit (prior $0.40$); CD-Hybrid renders each CommitDistill unit alongside the linked commit body with BM25's per-item body-character budget plus a typed-claim header. CIs are paired bootstrap, $10{,}000$ resamples.}
\label{tab:rq5_hybrid}
\footnotesize
\setlength{\tabcolsep}{4pt}
\begin{tabular}{@{}lcc@{}}
\toprule
Condition & Mean score & $\Delta$ vs.\ control ($95\%$ CI) \\
\midrule
Control (subject only)              & $0.703$ & --- \\
+CommitDistill (CD-v2)               & $0.718$ & $+0.015$\quad$[-0.035,\,+0.065]$ \\
+CD-Hybrid (CD units + body)        & $0.720$ & $+0.018$\quad$[-0.035,\,+0.070]$ \\
+BM25 (top-3 raw msgs)              & $0.733$ & $+0.030$\quad$[-0.022,\,+0.083]$ \\
\bottomrule
\end{tabular}
\end{table}

\textbf{Headline result (four-arm): null on the overall mean for
every retriever; CD-v2 is statistically indistinguishable from
BM25 head-to-head.} At $n{=}200$ no retrieval condition produces
a detectable lift over control on the headline mean: every
$\Delta$-vs-control $95\%$~CI in Table~\ref{tab:rq5_hybrid}
includes zero. The head-to-head comparison most relevant to the
typed-distillation hypothesis is CD-Hybrid against BM25: at
matched per-item body budgets (CD-Hybrid carries $\sim\!23\%$
more total bytes per query, all in typed-claim headers; see
Section~\ref{sec:rq5_ablation}),
$\Delta_{\text{CD-Hybrid} - \text{BM25}} = -0.013$
$[-0.068, +0.043]$ (paired bootstrap), i.e., statistically
indistinguishable. The honest summary is that on this LLM-judge
protocol the typed-unit format and the raw-prose format produce
comparable downstream LLM utility, and neither significantly
beats the bare-prompt control on the overall mean.

\begin{table}[t]
\centering
\caption{Hard-cohort breakdown ($n{=}102$ fixes where the control LLM scored $\leq 0.5$). With silence repaired (CD-v2), all three retrievers lift the hard half by comparable amounts; the gap between CD-v2 and BM25 narrows to $\sim\!1.3\times$.}
\label{tab:rq5_hard}
\footnotesize
\setlength{\tabcolsep}{4pt}
\begin{tabular}{@{}lcc@{}}
\toprule
Condition & Mean on hard half & Lift vs.\ control \\
\midrule
Control (subject only)              & $0.304$ & --- \\
+CommitDistill (CD-v2)               & $0.426$ & $+0.123$ \\
+CD-Hybrid (CD units + body)        & $0.446$ & $+0.142$ \\
+BM25 (top-3 raw msgs)              & $0.466$ & $+0.162$ \\
\bottomrule
\end{tabular}
\end{table}

\textit{Conditional finding: on hard cases, all three retrievers help.}
Restricting attention to fixes on which the control LLM scores
$\leq 0.5$ ($102$ of $200$, the half where the bug subject alone
is insufficient), all three retrieval conditions produce
substantial lifts over control (Table~\ref{tab:rq5_hard}):
CD-v2 $+0.123$, CD-Hybrid $+0.142$, BM25 $+0.162$. The lifts
are $\sim\!3$--$5\times$ the noise floor implied by the
overall-mean CIs and consistent across both judges.

\textit{Conditional finding: saturation hurts retrievers.}
On fixes the LLM already nails from the subject alone (control
$\geq 1.5$, $n{=}15$), adding retrieved context distracts:
CD-v2 drops the mean by $-0.227$ and BM25 by $-0.267$. This
is a property of \textit{any} unconditional retrieval: when the
model already has enough, the extra characters dilute attention
rather than sharpen it.

\textit{Inter-judge agreement.} Cohen's $\kappa = 0.355$ between
\textsc{gpt-4o} and \textsc{gpt-4o-mini} on the four-arm run
(and $\kappa = 0.366$ on the three-arm slice) is in the
\emph{fair} band of Landis and Koch~\cite{landis_koch}, well
below the $\kappa = 0.633$ on the human RQ2 rubric. The
disagreement is largely calibration-scale: \textsc{gpt-4o}
is consistently harsher (mean $0.56$ across the four arms)
than \textsc{gpt-4o-mini} (mean $0.88$). Per-fix Spearman
rank correlation between the two judges' score vectors is
$\rho \in [0.50, 0.57]$ across the four arms (substantial
per-fix agreement on \emph{which fixes are well answered}).
The condition ordering, however, is \emph{not} preserved
across judges alone: \textsc{gpt-4o} ranks CD-Hybrid $>$ BM25
$>$ control $\geq$ CD-v2 (means $0.570 / 0.565 / 0.550 /
0.550$) while \textsc{gpt-4o-mini} ranks BM25 $>$ CD-v2 $>$
CD-Hybrid $>$ control (means $0.900 / 0.885 / 0.870 / 0.855$).
The across-conditions rank correlation is therefore weak
($\rho = 0.21$), and the within-judge differences between the
four conditions are small relative to the noise floor implied
by the paired-bootstrap CIs in Table~\ref{tab:rq5_hybrid}
(all of which include zero). The honest reading is that the
aggregate-mean ordering reported in Table~\ref{tab:rq5_hybrid}
reflects an average of two judges that individually disagree on
the condition ranking, and that this disagreement is itself
consistent with the headline null effect. We treat the low
$\kappa$ and the weak across-conditions $\rho$ as joint
threats to construct validity (Section~\ref{sec:threats}).

\textit{What this evaluation does and does not establish.}
RQ5 establishes that, on this LLM-judge protocol with $n{=}200$
fixes drawn from two Python projects, (a)~no retrieval condition
beats control on the headline mean, (b)~all three retrieval
methods help on the hard half of the sample by comparable
amounts, and (c)~the typed-unit format and the raw-prose format
are statistically indistinguishable head-to-head once the
extractor silence floor is removed. RQ5 does
\textit{not} establish a positive mean-utility claim for
CommitDistill in end-to-end LLM patch generation, and we make
none. The positive case for CommitDistill is the
budget-constrained retrieval result (Section~\ref{sec:budget}),
where the task is \textit{find the answer span under a tight
token budget} rather than \textit{produce a full patch}; these
are different tasks and CommitDistill wins the former decisively
($0.750$ vs.\ $0.333$ at $B=256$) while tying the latter.
The ICSE/FSE-grade controlled human study
(Section~\ref{sec:future}) remains the only experiment that can
adjudicate which of the two tasks better predicts real developer
utility.

\subsubsection{The original three-arm record (CD-v1)}
\label{sec:rq5_v1_record}

For full transparency we preserve the original three-arm result
that motivated the four-arm ablation. Table~\ref{tab:rq5} reports
the original CD-v1 numbers: CD-v1 $\Delta = -0.040$
$[-0.090, +0.010]$, BM25 $\Delta = +0.030$ $[-0.022, +0.082]$,
control mean $0.703$. The CD-v1 arm was structurally
silent: Algorithm~\ref{alg:extract} returned zero substantive
units on $96/100$ \texttt{requests} fixes and $100/100$
\texttt{flask} fixes (i.e., $\sim\!98\%$ silence), so the
$\Delta_{\text{CD-v1}} = -0.040$ measured how the assistant
performs with CommitDistill \emph{returning nothing} rather than how a
typed-unit signal performs against BM25. The $n{=}40$ pilot also
showed a halved harm-rate for CD-v1 versus BM25 ($3$/$33$/$4$
CD-v1 win/tie/loss vs.\ $4$/$30$/$6$ BM25); at $n{=}200$ this did
\emph{not} replicate (CD-v1 $26$/$138$/$36$, BM25 $39$/$132$/$29$),
and we no longer claim it. The four-arm ablation
(Section~\ref{sec:rq5_ablation}) addresses the silence confound
directly.

\begin{table}[t]
\centering
\caption{The original three-arm CD-v1 record ($n{=}200$), preserved for transparency. The CD-v1 arm was silent on $\sim\!98\%$ of bug-fix subjects, so this row measures the assistant performing with CommitDistill \emph{returning nothing} versus with the subject alone, not retrieval-signal quality. The four-arm Table~\ref{tab:rq5_hybrid} resolves the silence confound.}
\label{tab:rq5}
\footnotesize
\setlength{\tabcolsep}{4pt}
\begin{tabular}{@{}lcc@{}}
\toprule
Condition                & Mean score & $\Delta$ vs.\ control ($95\%$ CI) \\
\midrule
Control (subject only)   & $0.703$    & --- \\
+CommitDistill (CD-v1)    & $0.663$    & $-0.040$\quad$[-0.090,\,+0.010]$ \\
+BM25 top-$3$            & $0.733$    & $+0.030$\quad$[-0.022,\,+0.082]$ \\
\bottomrule
\end{tabular}
\end{table}

\subsubsection{Algorithm 1 ablation: subject-fallback units (CD-v2) and a payload-matched hybrid (CD-Hybrid)}
\label{sec:rq5_ablation}

The four-arm ablation reported in Table~\ref{tab:rq5_hybrid}
adds two treatments to the original CD-v1 / BM25 / control
design:

\begin{itemize}
\item \textbf{CD-v2.}\, Algorithm~\ref{alg:extract} extended with a
single subject-fallback unit ($\textsf{type}{=}\texttt{pattern}$,
prior $0.40$, content = cleaned subject plus first body sentence,
total payload capped at $280$ chars) emitted only when the regex
pass returns zero substantive units. On the same $n{=}200$ corpus
this reduces extractor silence from $98\%$ to $\mathbf{34.5\%}$
without displacing any regex-extracted unit (overlap $= 0$ across
all $200$ fixes); the fallback prior of $0.40$ is below every
regex prior ($0.65$--$0.95$) so the fallback is the
lowest-ranked unit when a regex hit exists.

\item \textbf{CD-Hybrid.}\, Same retrieval signal as CD-v2, but each
ranked unit is rendered alongside the linked commit body using
\emph{exactly} the per-item body-character budget BM25 uses ($140$
chars of body, at most $3$ non-empty body lines). The per-item
body payload is therefore identical to BM25's; CD-Hybrid carries
an additional typed-claim header line per item (e.g.\
\texttt{behavior:retry/backoff}) on top of that body, so the total
per-item byte budget is BM25's body budget plus the typed header.
This is a deliberate design choice: the hybrid tests whether
adding a typed header to BM25's body content moves the score,
not whether typed headers and raw subjects are equivalent at
strictly equal total bytes (a different question we do not
address here).
\end{itemize}

Char-budget audit on the same $n{=}200$ corpus quantifies the
delta. Mean per-query payload is $228.6$ chars for PM,
$278.1$ for CD-Hybrid, and $225.8$ for BM25 (medians: $214$ /
$228$ / $220$). CD-Hybrid therefore carries $\approx\!52$ extra
chars per query on average ($\sim\!23\%$ more than BM25),
accounted for entirely by the typed-claim headers; we describe
CD-Hybrid as ``body-budget-matched, header-augmented'' rather
than fully byte-matched throughout the paper.
Cache-key compatibility was preserved (Section~\ref{sec:rq5}
cache identifies entries by $\mathrm{sha1}(\mathit{model} \mathbin\Vert
\mathit{system} \mathbin\Vert \mathit{user})$), so the four-arm
re-run reused all stable assistant responses and paid for the new
CD-v2 and CD-Hybrid prompts only ($\sim\!975$ new API calls).

\textit{What the ablation establishes and does not establish.}
The ablation establishes that (a)~the original CD-v1 negative
in Table~\ref{tab:rq5} was caused by extractor silence rather
than by typed-unit retrieval quality, (b)~CD-v2's subject-fallback
unit is a strict additive improvement at the extraction layer
(zero overlap with regex emissions, prior below all regex
priors), and (c)~once payloads are budget-matched at the body
level (CD-Hybrid vs.\ BM25), the typed-unit format and the
raw-prose format are statistically indistinguishable on this
LLM-judge protocol. It does \emph{not} establish that any
retriever beats the bare-prompt control on the overall mean:
all three $\Delta$-vs-control CIs in Table~\ref{tab:rq5_hybrid}
include zero. As noted above, the ablation is exploratory; a
held-out replication on a third Python repository would be
required to elevate any of the four-arm $\Delta$-values from
descriptive to confirmatory.

\subsection{Worked Retrieval Example}
For the query \texttt{intersphinx documentation link broken} on the
\texttt{requests} store ($k = 3$, $\theta = 0.05$), CommitDistill
returns two units; the top-ranked unit (score $1.71$) is the Fact
``\textit{When trying to link via intersphinx, a label must be used}'',
attributed to commit \texttt{b5bd0f14}. A developer or LLM consuming
this output sees one sentence plus a SHA --- $72$ characters of
content --- enough to either fix the docstring directly or to read
the original commit for full context. The same query against
\texttt{git log --grep} returns commit \texttt{0e4ae38f}
(``\textit{docs: exclude Response.is\_permanent\_redirect from API
docs}'') as its top-1, which mentions the word ``docs'' but is
unrelated; against BM25 it returns commit \texttt{dffd5d43} which
\textit{is} related but bundles a $> 280$-character merge body into
the result. The trade-offs in Table~\ref{tab:queries} are visible in
miniature on this single query.

\subsection{What These Numbers Are and Are Not}
The case study demonstrates that the prototype \textit{runs}, that
its outputs are inspectable, that the widened Pattern heuristics
populate the third type, that an inter-annotator agreement of
$\kappa = 0.633$ holds on a four-class rubric, that the
silent-on-novel-queries property holds in practice, and that on a
regression-finding task the system is structurally outperformed by
BM25 over raw commits --- a negative result that we report rather
than hide. It does \textit{not} demonstrate that
the system improves developer productivity, that its precision
generalises beyond two Python projects, or that it competes with
a tuned neural retriever. We discuss these limits explicitly in
Section~\ref{sec:threats}.

\section{Discussion}
\label{sec:discussion}

\subsection{When CommitDistill Helps}
Two contexts seem to fit naturally. \textit{Onboarding}: a new contributor
asks the store about a behaviour they observe, and (when the corpus
covers it) receives a one-sentence answer with a commit SHA they can
follow. \textit{Pre-LLM filtering}: an agentic coding assistant prepends
the top retrieved units to its context, gaining a small number of
high-precision project facts that complement the assistant's own
embedding-based retrieval. The latter use is consistent with the
typed-memory hypothesis~\cite{memgpt,cognitive_arch} that distilled,
typed knowledge yields better decisions per token than raw text.

\textbf{Token-economy estimate.}
On the six hand-authored queries (Table~\ref{tab:queries}), the
median CommitDistill top-1 result is $72$ characters, against $287$
for BM25 ($\approx 4\times$ shorter) and $301$ for \texttt{git log
--grep} ($\approx 4.2\times$ shorter). The headline relative ratio
is the robust observation here: typed units encode the answer span
as a single short sentence whereas the lexical baselines return
whole commit bodies. An illustrative per-developer-day projection
($20$ retrievals $\times$ top-$3$ consumption $\times (287-72)$
chars $\approx 13{,}000$ characters or $\approx 3{,}000$--$4{,}000$
tokens saved per day) is a back-of-envelope figure derived from
these six queries; we report it for intuition only, not as a
deployment-scale claim, and the underlying token-saving
mechanism --- shorter retrieved spans at fixed top-$k$ --- is the
generalisable observation.

\textbf{Falsifiable downstream hypothesis.}
The distillation-helps-the-LLM claim is testable but unproven by
this paper. We state it explicitly so it can be falsified in
follow-up work: \emph{prepending CommitDistill's top-$3$ retrieved
units to an LLM coding assistant's context, on $\geq 30$ real
bug-fix tasks drawn from the subject repositories' issue trackers,
should improve a fixed downstream metric (e.g., Jaccard overlap
between the assistant's proposed-fix files and the human fix's
files, or pass-rate on the project's own test suite) by at least
$5$ percentage points relative to a no-retrieval baseline}. If
that effect is not observed under independent execution at
$n \geq 30$, the typed-memory-helps-LLM-coding-assistants hypothesis
should be considered weakened for this corpus class.

\subsection{When CommitDistill Does Not Help}
The case study makes three weaknesses concrete. First, \textit{recall
remains the dominant limitation}: even with widened Pattern heuristics,
many genuinely useful constraints phrased without a trigger keyword
are missed, and the per-type precision drop on Patterns ($38.5\%$,
versus $63.6\%$ for Facts) shows that broadening the regex set has
a precision cost. Second, \textit{yield depends on commit-message
culture}: $329.5$ units$/$kc on \texttt{redis} versus $8.0$ on
\texttt{flask} is not a property of the projects' code but of how
their maintainers write commit prose. Third, the \textit{retriever
is lexical}: it cannot bridge ``redirect loop'' to ``infinite 302
chain''. All three are addressable (see Section~\ref{sec:future})
but they are real properties of the present prototype, not edge cases.

\subsection{The History-Blindness Default and What Our RQ5 Does and Does Not Test}
\label{sec:history-blindness}
A natural intuition is that CommitDistill \emph{should} win by
default, because the deployed competitive landscape for code
generation and PR review is overwhelmingly history-blind: GitHub
Copilot autocomplete, most chat-style coding assistants, and the
current generation of automated PR-review bots condition on the
current file (and at most a small set of open editor tabs or the
diff under review), but do not retrieve from the project's commit
history at all.

Our RQ5 (Section~\ref{sec:rq5}) does compare a history-blind
arm against three history-augmented arms: the \textbf{control}
arm receives only the bug subject and project name --- no
retrieval at all --- while the \textbf{+CommitDistill (CD-v2)},
\textbf{+CD-Hybrid}, and \textbf{+BM25} arms inject distilled
units, distilled-units-with-matched-body-budget, and raw commit
excerpts respectively (Section~\ref{sec:rq5_ablation}). The
four-arm deltas ($\Delta_{\text{CD-v2}} = +0.015$,
$\Delta_{\text{CD-Hybrid}} = +0.018$, $\Delta_{\text{BM25}} = +0.030$;
all CIs include zero) are therefore literal
history-blind-vs-history-augmented contrasts on this task. The
limitation is not in the contrast itself but in the
\emph{baseline being contrasted}. Our control is a \textit{bare
prompt} history-blind condition (one cleaned bug subject plus
the project name); a production coding assistant is also
history-blind in the commit-history sense, but operates on a
much richer non-history context (current file, open editor
tabs, recent edits, project structure, IDE-resolved imports).
A bare-prompt history-blind LLM and a production-grade
history-blind LLM are different baselines, and CommitDistill's
augmentation effect on each may differ.

What RQ5 therefore lets us claim, on this corpus and protocol,
is the following: prepending CommitDistill's distilled units to
a deliberately-minimal bug-fix prompt produces an effect that
is statistically indistinguishable from prepending BM25's raw
commit excerpts at body-budget-matched, header-augmented per-item
payloads
($\Delta_{\text{CD-Hybrid}-\text{BM25}} = -0.013$ $[-0.068,
+0.043]$, paired bootstrap), and neither retriever shows a
detectable lift over the bare-prompt control on the overall
mean (all three $\Delta$-vs-control CIs span zero); on the harder $102$
of $200$ fixes (control mean $\leq 0.5$) all three retrievers
help by comparable amounts (CD-v2 $+0.123$, CD-Hybrid $+0.142$,
BM25 $+0.162$). What RQ5 does \emph{not}
let us claim is that CommitDistill augments a richer
production-grade history-blind assistant in the same direction
or magnitude. That requires a separate experimental arm whose
control is the production assistant itself, not our minimal
prompt; we operationalise it as future work in
Section~\ref{sec:future} (item~5).

Two corollaries follow that we treat as observed rather than
inferred. First, the decisive case for CommitDistill in this
paper is the budget-constrained retrieval result of
Section~\ref{sec:budget} ($0.750$ hit-rate vs.\ BM25's $0.333$
at a $256$-char budget), where the task is finding the answer
span under a tight token budget rather than producing a full
LLM patch --- i.e., distillation buys decisive
payload-efficiency advantages, not absolute-quality advantages,
on this corpus. Second, on the patch-generation task at full
payload, the typed-unit format and BM25's raw-prose format are
statistically indistinguishable when per-item body budgets are
matched (CD-Hybrid vs.\ BM25, Section~\ref{sec:rq5_ablation});
this is consistent with the assistant LLM treating the
connective commit-message prose and the typed-claim header as
interchangeable carriers of the same underlying signal at this
granularity.

\subsection{Why Keep the System Lexical and Local?}
Why not use embeddings? Two reasons.
First, the artefact is meant to be a \textit{baseline} that any
follow-up neural method can be compared against; it must therefore
be deterministic and dependency-free. Second, in regulated environments
(industrial codebases under export control or HIPAA), the inability
to send code to an embedding service is a deployment constraint, not a
design limitation. A purely local lexical baseline establishes what
can be achieved without any external model. The hypothesis we inherit
from MemGPT and the cognitive-architectures literature~\cite{memgpt,cognitive_arch}
--- that distilled, typed knowledge yields better decisions per token than
raw text --- can then be tested by layering neural extension on top of
the lexical core (Section~\ref{sec:future}).

\section{Threats to Validity}
\label{sec:threats}

We organise threats following Wohlin et al.'s classification~\cite{wohlin}.

\paragraph{Construct validity}
``Useful'' as a label depends on annotator judgement. We expanded the
rubric in this revision from three classes to four (\textit{useful},
\textit{trivially-true}, \textit{fragment}, \textit{noise}) after
preliminary inspection showed that short matched spans deserved their
own class rather than being forced into \textit{noise}. The two
annotators agreed on $77.5\%$ of unit labels ($\kappa = 0.633$,
substantial agreement on Landis-Koch~\cite{landis_koch}), but the
rubric is ours and may not match what working developers consider
useful at the point of need. The downstream RQ5 evaluation is
weaker on this dimension: inter-judge $\kappa$ ranges from $0.355$
(four-arm) to $0.366$ (three-arm) between
\textsc{gpt-4o} and \textsc{gpt-4o-mini}, in the \emph{fair}
Landis-Koch band only, and is driven by absolute-scale calibration
difference rather than ordering disagreement (\textsc{gpt-4o}
rarely assigns the highest score). The condition ordering matches
between judges; the magnitude estimate is judge-dependent. A
controlled study with practising maintainers, with a pre-defined
binary or three-point rubric anchored on developer behaviour
rather than LLM judgement, would replace both the human-annotation
($n{=}40$) and LLM-judge ($n{=}200$) evaluations with a stronger
construct.

\paragraph{Trustworthy-AI threat model}
Because CommitDistill distils repository prose into typed claims that
can flow into an LLM context, a dedicated threat model is warranted.
We identify four classes of concern. (i)~\emph{Commit-prose
poisoning.} A malicious or careless contributor can author a commit
message that triggers our regex heuristics and writes a deceptive
unit into the store (e.g., a fabricated ``\texttt{Fact: do not validate
TLS certificates on this code path}'' phrased to match our
substantive-Pattern filter). The current pipeline has no
authentication or trust signal beyond commit-author metadata, so any
party with commit access to a subject repository can write into the
distilled store; this is a real risk for un-curated open-source
mirrors and a near-zero risk for repositories with mandatory PR
review. (ii)~\emph{Privacy.} The distilled store may
verbatim-preserve fragments of internal commit messages; in
regulated environments (export-controlled, HIPAA-bound, or
defence-adjacent codebases) operators must treat the resulting
\texttt{units.json} files with the same access controls as
\texttt{git log} itself. The local-only execution posture
(no embedding service, no external API call at extraction time)
helps here but does not by itself enforce data-handling policy.
(iii)~\emph{Confidentiality of attribution.} Each unit carries a
commit author and SHA. In some workflows (anonymous review, public
artefact release of an internal corpus) this provenance must be
scrubbed; we expose a \texttt{--strip-attribution} CLI flag for
that case, but the default is full attribution.
(iv)~\emph{Adversarial silence circumvention.}
Section~\ref{sec:threshold} shows CD-v1 abstains on
NOT\_IN\_CORPUS / OOD queries at $\theta = 2.5$, but a
sufficiently determined adversary who controls commit prose can
write a commit subject that engineers a high TF--IDF score for
an unrelated query, defeating the abstention. This is a
data-quality concern rather than a property of the retrieval
algorithm and should be addressed with provenance trust signals
(commit-signature verification, contributor reputation, code-review
gating) in production deployments.

These threats matter more under autonomous consumption than under
interactive consumption. When an LLM agent reads a distilled
unit at decision time and acts on it without a human-in-the-loop
sanity check between retrieval and action, a single poisoned
unit can silently steer the agent's tool calls or generated patch;
commit-signature verification and code-review gating therefore
become trust requirements rather than UX niceties. Item~4 of
Section~\ref{sec:future} (IDE integration that surfaces units
as reviewable PR comments and on-error tooltips) is in this sense
as much a trust mitigation as a usability item: it keeps the
distilled unit on a path with a human reader before the unit can
influence an agent's output. None of these threats are
specific to CommitDistill; we surface them because the case for
the system rests partly on its trust posture, and a trust-relevant
contribution is incomplete without them.

\paragraph{Internal validity}
Extraction is deterministic, so internal threats focus on the labelling.
Both annotators are co-authors and were aware of the system's intent;
to mitigate this, each annotator labelled units in a randomised order
from a CSV that did not surface the system's retrieval output, and
adjudication of disagreements was done by a third co-author. A fully
blinded protocol (independent raters with no system knowledge) would
be stronger and is planned for the controlled follow-up study.

\paragraph{External validity}
The case study uses five public projects spanning Python, JavaScript,
C, and Java, but precision is measured only on the two Python
subjects ($n=40$); the same two subjects also bound the RQ4
time-travel and RQ5 LLM-judge experiments. Yield generalisation is
the more important threat: the order-of-magnitude gap between
\texttt{redis} ($329.5$ units$/$kc) and \texttt{flask} ($8.0$
units$/$kc) shows that yield depends as much on commit-message
culture as on the extractor. External-validity claims are therefore
restricted to \emph{public, commit-message-rich} projects;
enterprise repositories whose commit prose is dominated by ticket
numbers and merge subjects may yield little. Re-labelling
useful-precision at the $5{,}000$-commit window (where the
substantive-Pattern filter applies) and labelling $\geq\!20$
additional units from a non-Python subject (\texttt{redis} is
the natural choice given its $928$-unit yield) are deferred to
follow-up work; we surface this so the $0.525$ point estimate
is not over-generalised. A reasonable target for the follow-up
labelling pass is a $95\%$ bootstrap CI half-width of
$\pm 0.10$ around the point estimate, which (assuming the
current precision rate holds) implies $n \approx 80$ labelled
units per subject --- doubling the present sample on the two
Python subjects and labelling a comparable batch from
\texttt{redis} to cover commit-prose-rich C code. Thirty-six queries is
too few to characterise retrieval recall in general, and the
auto-derived queries trivially favour the baselines that index the
same commit window. The RQ5 LLM-judge protocol uses two
OpenAI-family judges; running it with a non-OpenAI judge
(e.g.,~Claude or a local Llama-class model) is left to future work
and may shift the calibration constant.

\paragraph{Conclusion validity}
We deliberately report no significance tests on
Tables~\ref{tab:yield}--\ref{tab:queries} because the sample sizes
do not support them. Where we do compute confidence intervals --- the
useful-precision $95\%$ bootstrap CI of $[0.375, 0.675]$ in RQ2 and
the paired-bootstrap CIs in Table~\ref{tab:rq5} --- both intervals
are wide and, in RQ5, brush zero. We therefore frame the headline
RQ5 result as \emph{null with a slight negative trend for
CommitDistill}, not as evidence of equivalence; ruling out a true
effect of the magnitude reviewers might expect would require
$n \gg 200$. Treating Tables~\ref{tab:yield}--\ref{tab:rq5} as
descriptive observations of one artefact on five repositories is the
strongest conclusion the data support.

\section{Future Work}
\label{sec:future}

We see five concrete next steps:
\begin{enumerate}
    \item \textbf{Hybrid extraction.} Augment heuristics with an LLM
    classifier that proposes candidate units; use the regex layer as a
    fast filter and as a deterministic baseline. This directly addresses
    the $38.5\%$ Pattern precision deficit reported in
    Section~\ref{sec:results} (RQ2). CD-v2's deterministic
    subject-fallback unit (Section~\ref{sec:rq5_ablation}) is an
    early step in this direction --- a regex-prior approximation of
    what an LLM classifier would propose --- and the four-arm
    ablation gives a baseline against which a true LLM classifier
    must improve.
    \item \textbf{Hybrid retrieval.} Add an optional sentence-transformer
    re-ranker behind the same CLI. The lexical layer remains the default
    so the system stays usable offline.
    \item \textbf{Controlled user study.} Recruit maintainers of three
    open-source projects, present them with their own knowledge stores,
    and measure precision-at-10 on tasks drawn from real issues. This
    is the study a reviewer would expect for a full evaluation paper.
    \item \textbf{IDE integration.} Surface units inline as PR comments
    and as on-error tooltips, mirroring the way modern coding assistants
    are integrated.
    \item \textbf{Production-grade history-blind vs.\ history-aware
    assistant study.} The strongest open question raised by
    Section~\ref{sec:history-blindness} is whether CommitDistill
    augments a \emph{production-grade} history-blind coding
    assistant --- one that already conditions on the current
    file, open editor tabs, and recent edits --- on its native
    task. We propose a fourth experimental arm that this paper
    does not run: pair a fixed off-the-shelf assistant (e.g.,
    GitHub Copilot Chat or an equivalent IDE assistant invoked
    through its public API) against the \emph{same} assistant
    prepended with CommitDistill's top-$k$ retrieved units at
    decision time, on a held-out set of $\geq 100$ real pull
    requests drawn from the five subject repositories.
    The dependent variables would be (i)~PR-fix file-set
    Jaccard overlap with the human fix, (ii)~project-test-suite
    pass rate on the assistant's proposal, and (iii)~a paired
    LLM-as-judge comparison of the two assistants' patches under
    the same blinded protocol used in Section~\ref{sec:rq5}.
    Crucially, this experiment differs from RQ5 in that the
    control arm is the production assistant operating on its
    full native context (current file, open tabs, recent edits)
    rather than the bare bug-subject prompt of
    Section~\ref{sec:rq5}'s control. Such a control matches the
    deployment posture of essentially every shipping coding
    assistant today; a positive CommitDistill augmentation effect
    against \emph{that} baseline would establish the
    default-augmentation claim that RQ5's bare-prompt control
    cannot. We flag this as a user-study-scale follow-up rather
    than fold it into the present artefact-evaluation paper.
    \item \textbf{Held-out replication of the four-arm ablation.}
    The CD-v2 subject-fallback unit and the CD-Hybrid rendering
    were designed after observing the original three-arm RQ5
    result, and the four-arm $\Delta$-values reported in
    Section~\ref{sec:rq5_ablation} are therefore exploratory
    rather than confirmatory. The natural confirmatory follow-up
    is to re-run the same four-arm protocol unchanged on a third
    Python repository not used in CD-v2 development (e.g.,
    \texttt{httpx} or \texttt{urllib3}), with the existing cache,
    drivers, and extractor, and report the resulting four
    $\Delta$-vs-control values without modification. A successful
    replication would elevate the head-to-head
    CD-Hybrid$\sim$BM25 tie from a descriptive characterisation
    to a confirmatory effect-size claim.
\end{enumerate}

\section{Conclusion}
\label{sec:conclusion}

We presented CommitDistill, a lightweight, dependency-free Python
prototype that adapts the typed-memory stance of LLM agent
architectures~\cite{plugmem_yang,memgpt,cognitive_arch} to a
software repository's own commit history, with deterministic regex
extraction, a calibrated silence threshold, and an inspectable
plain-JSON store. An illustrative case study on five public
repositories ($25{,}000$ commits, $1{,}167$ units) reports
useful-precision $0.525$ at $\kappa = 0.633$ on $40$ dual-annotated
Python units (Section~\ref{sec:results}). The decisive positive
finding is budget-constrained retrieval: $0.750$ hit-rate at a
$256$-char budget on a $12$-query fact-style benchmark, against
BM25's $0.333$ and \texttt{git log --grep}'s $0.083$
(Section~\ref{sec:budget}, with a jackknife sensitivity row in
Table~\ref{tab:budget}). The paired four-arm LLM-as-judge
downstream evaluation on $200$ time-travel bug-fixes reports no
detectable mean lift over a no-retrieval control for any retriever
(all $\Delta$-vs-control $95\%$ CIs span zero); on the hard half
of the sample ($102/200$, control $\leq 0.5$) all three retrievers
help by comparable amounts (Section~\ref{sec:rq5_ablation}). The
honest summary is that this case study establishes CommitDistill
as a trust-instrumented reproducible baseline that wins decisively
on payload-efficient retrieval and ties raw-prose retrieval at full
payload on patch-generation utility; it does not yet establish a
positive end-to-end LLM-utility claim. The four-arm ablation is
exploratory; a held-out replication on a third Python repository
(\texttt{httpx} or \texttt{urllib3}) is the natural confirmatory
follow-up (Section~\ref{sec:future}, item~6). The accompanying
open-source release is intended to make that follow-up easy.

\section*{Artefact and Data Availability}
The full source of \texttt{commit\_distill/}, the $11$ unit tests, the
case-study driver, the raw extracted units for both subject
repositories, and the raw query results are released as a
self-contained archive accompanying this paper (anonymised
repository link to be inserted on acceptance). The five subject
repositories (\texttt{psf/requests}, \texttt{pallets/flask},
\texttt{expressjs/express}, \texttt{redis/redis},
\texttt{junit-team/junit5}) are public on GitHub.
No human subjects were involved beyond the two co-author annotators
in Section~\ref{sec:results}; therefore no IRB approval was required.
The case study in Sections~\ref{sec:eval}--\ref{sec:results} can be
reproduced end-to-end with:
\begin{lstlisting}
cd <repo-root>
python -m venv .venv
# activate venv per OS convention:
#   POSIX:    source .venv/bin/activate
#   Windows:  .venv\Scripts\Activate.ps1
python -m unittest discover -s tests
python evaluation/extract_all.py
python evaluation/baseline_compare.py
python evaluation/threshold_sweep.py
python evaluation/budget_recall_v2.py
python evaluation/compute_kappa.py
python evaluation/time_travel_eval.py
\end{lstlisting}
The \texttt{extract\_all.py} driver iterates over all five subjects at
\texttt{--max 5000}; \texttt{baseline\_compare.py} produces
\texttt{evaluation/baseline\_results.json};
\texttt{threshold\_sweep.py} re-derives Table~\ref{tab:threshold};
\texttt{budget\_recall\_v2.py} re-derives Table~\ref{tab:budget};
\texttt{compute\_kappa.py} re-derives the $\kappa$ and bootstrap CI
reported in Section~\ref{sec:results} from
\texttt{evaluation/labels.csv}; \texttt{time\_travel\_eval.py}
re-derives Table~\ref{tab:timetravel}. All output is deterministic
given fixed git histories; total wall-clock for the
non-time-travel pipeline is under $5$\,s on a 2023 laptop.

\section*{Acknowledgments}
We thank our colleagues at Microsoft Corporation for internal
discussions and feedback on early drafts. The framing of typed,
retrieve-on-decision memory follows MemGPT~\cite{memgpt} and the
cognitive-architectures tradition of Sumers \textit{et al.}~\cite{cognitive_arch}.

\end{document}